\documentclass[sigconf]{acmart}
\pdfoutput=1
\usepackage[utf8]{inputenc}
\usepackage{color}

\usepackage{microtype}
\usepackage{balance}
\usepackage{todonotes}
\usepackage{siunitx}
\usepackage{listings}
\usepackage{cleveref}
\usepackage{pgfplots}
\usepackage{booktabs}

\usetikzlibrary{pgfplots.dateplot}

\newcommand{\MYURL}[3]{\href{#1}{#2}\footnote{\url{#1}, accessed #3.}}
\newcommand{\MYURLp}[4]{\href{#1}{#2}#4\footnote{\url{#1}, accessed #3.}}  \newcommand{\swhid}[1]{\href{https://archive.softwareheritage.org/#1}{{\small\ttfamily #1}}}

\definecolor{guixorange1}{RGB}{243,154,38}  \definecolor{guixblue1}{RGB}{38,109,131} 

\definecolor{color-cb1}{HTML}{8dd3c7}
\definecolor{color-cb2}{HTML}{ffffb3}
\definecolor{color-cb3}{HTML}{bebada}
\definecolor{color-cb4}{HTML}{fb8072}
\definecolor{color-cb5}{HTML}{80b1d3}
\definecolor{color-cb6}{HTML}{fdb462}
\definecolor{color-cb7}{HTML}{b3de69}

\pgfplotscreateplotcyclelist{cb colors}{
  color-cb1!35!black,fill=color-cb1\\
  color-cb2!35!black,fill=color-cb2\\
  color-cb3!35!black,fill=color-cb3\\
  color-cb4!35!black,fill=color-cb4\\
  color-cb5!35!black,fill=color-cb5\\
  color-cb6!35!black,fill=color-cb6\\
  color-cb7!35!black,fill=color-cb7\\
}

\copyrightyear{2024}
\acmYear{2024}
\setcopyright{rightsretained}
\acmConference[ACM REP '24]{ACM Conference on Reproducibility and
  Replicability}{June 18--20, 2024}{Rennes, France}
\acmBooktitle{ACM Conference on Reproducibility and Replicability (ACM REP
  '24), June 18--20, 2024, Rennes, France}\acmDOI{10.1145/3641525.3663622}
\acmISBN{979-8-4007-0530-4/24/06}

\title{Source Code Archiving
  to~the~Rescue~of~Reproducible~Deployment}

\author{Ludovic Courtès}
\email{ludovic.courtes@inria.fr}
\orcid{0000-0003-4472-154X}
\affiliation{
  \institution{Inria}
  \city{Bordeaux}
  \country{France}}

\author{Timothy Sample}
\email{samplet@ngyro.com}
\orcid{0009-0007-1813-4073}
\affiliation{
  \city{Saskatoon}
  \country{Canada}
}

\author{Simon Tournier}
\email{simon.tournier@inserm.fr}
\orcid{0000-0002-2639-818X}
\affiliation{
  \institution{Université Paris Cité}
  \city{Paris}
  \country{France}}

\author{Stefano Zacchiroli}
\email{stefano.zacchiroli@telecom-paris.fr}
\orcid{0000-0002-4576-136X}
\affiliation{\institution{LTCI, Télécom Paris, Institut Polytechnique de Paris}
  \city{Palaiseau}
  \country{France}
}

\begin{document}
\begin{abstract}
The ability to \emph{verify} research results and to \emph{experiment}
with methodologies are core tenets of science.  As research results are
increasingly the outcome of computational processes, software plays a
central role.  GNU~Guix is a software deployment tool that supports
\emph{reproducible} software deployment, making it a foundation for
computational research workflows.  To achieve reproducibility, we must
first ensure the source code of software packages Guix deploys remains
available.

We describe our work connecting Guix with Software Heritage, the
universal source code archive, making Guix the first free software
distribution and tool backed by a stable archive.  Our contribution is
twofold: we explain the rationale and present the design and
implementation we came up with; second, we report on the archival
coverage for package source code with data collected over five years and
discuss remaining challenges.
\end{abstract}


\maketitle

\section{Introduction}
\label{sec:intro}

Now that software is an integral part of scientific experimental
workflows, it must be held to the same standards of other scientific
workflows: computational workflows must be transparent and reproducible.
While such a statement is becoming consensual among scholars, its
implications are often less understood: source code must be publicly
available, with a license that grants the right to use it, to study it,
to modify it, and to share those modifications.  UNESCO's Recommendation
on Open Science~\cite{unesco2021:openscience} further states:
\begin{quote}
In the context of open science, when open source code is a component of
a research process, enabling reuse and replication generally requires
that it be accompanied with open data and open specifications of the
environment required to compile and run it.
\end{quote}

That last part---the \textit{specifications of the environment}---is
often overlooked or dismissed: researchers often resort to either
imprecise natural-language build instructions or large binary images
(``containers'' or virtual machines) that let others run the software,
but typically prevent them from experimenting with the
code~\cite{vallet2022:toward}.

Guix is a software deployment tool developed by a large community since
2012 and which is seeing growing adoption~\cite{courtes2013:functional,
  courtes2022:secure}.
It can be used as a ``package manager'' such as those found on GNU/Linux
distributions---\texttt{apt}, \texttt{dnf}, \texttt{pip}, etc.---or as a
standalone system---just like Debian, Fedora, etc.  It builds upon the
\emph{functional software deployment} model pioneered by
Nix~\cite{dolstra2004nix} and \emph{reproducible
builds}~\cite{lamb2021:reproducible}, making it a tool of choice when
deploying software for reproducible research workflows.

Instead of capturing a list of package names and versions, Guix lets
users capture the entire graph of package definitions, which includes,
beyond version numbers, all the information required to build the
packages, including how to fetch the source code and what its
cryptographic hash must be, configuration options, and
dependencies---\emph{recursively}.  Once they have captured this
information, users can run the \texttt{guix time-machine} command to
re-deploy \textit{the exact same software environment}, bit for bit, on
a different machine or at a different point in
time~\cite{vallet2022:toward}.

This, of course, can only be true if one necessary condition holds: that
package source code is available.  Indeed, when the source code of a
package disappears, Guix (unsurprisingly) cannot build that package
anymore and thus becomes unable to redeploy the software environment.
The entire foundation for reproducible deployment collapses if the
source of one or more of the packages in our environment disappears.  This
scenario is far from unrealistic: source code hosting sites come and go,
even those backed by large companies deemed ``too big to
fail''~\cite{escamilla2022:github}.

With the mission to save and archive all publicly-available source code,
Software Heritage~\cite{dicosmo2017swh} (SWH for short) has the potential to
let us fill this gap.

\paragraph{Contributions and paper structure}
In the vast field of reproducible research, we focus exclusively on
software deployment and under the assumption that we are only dealing
with free and open source software, in line with the open science
recommendations mentioned above.
The following sections describe what it meant to
``connect'' Software Heritage and Guix, and the roadblocks we had to overcome.
\Cref{sec:implementation} describes our first contribution: the
design and implementation of our bridge between Guix and SWH, including
novel tools developed to address sub-problems.
\Cref{sec:evaluation} provides an evaluation of the effectiveness
of our solution, looking at 5 years worth of package source code
referred to by Guix---our second contribution.
Lastly, we describe related work and conclude on
challenges that remain to be addressed.

\paragraph{Reproducibility package}
A full reproducibility package for this work is available.
See \Cref{sec:data-availability} for details.

\section{Background}
\label{sec:background}

To understand what we are trying to achieve, let us first describe the
two components at play: Guix package definitions on one hand, and the
Software Heritage archive on the other hand.

\subsection{Guix Package Definitions}
\label{sec:background:guix}

Guix is a software deployment tool that stands alone: it can only deploy
software packages that have been \emph{defined} in its own package
collection.  To date, Guix itself provides almost \num{30 000} packages,
making it one of the five largest free software distributions according
to
\MYURLp{https://repology.org}{Repology}{2024-01-18}{.}
Each package definition specifies metadata, instructions on how to build the
package, and references to dependencies (which are themselves other Guix
packages).  Like Nix and unlike Debian or Fedora, Guix at its core is a
``source-based'' deployment tool that builds software from source; it
can optionally download pre-built binaries as a substitute for local
compilation~\cite{dolstra2004nix, courtes2022:secure}.

\begin{figure}
  \begin{lstlisting}[language=Scheme]
(define-public python
  (package
    (name "python")
    (version "3.10.7")
    (source
     (origin
       (method url-fetch)
       (uri (string-append "https://www.python.org/ftp/python/"
                           version "/Python-" version ".tar.xz"))
       (sha256
        (base32
         "0j6wvh2ad5jjq5n7sjmj1k66mh6lipabavchc3rb4vsinwaq9vbf"))))
    ;; various fields omitted
    (license license:psfl)))

(define-public python-scikit-learn
  (package
    (name "python-scikit-learn")
    (version "1.3.2")
    (source
     (origin
       (method git-fetch)
       (uri (git-reference
             (url "https://github.com/scikit-learn/scikit-learn")
             (commit version)))
       (sha256
        (base32
         "1hr024vcilbjwlwn32ppadri0ypnzjmkfxhkkw8gih0qjvcvjbs7"))))
    ;; various fields omitted
    (license license:bsd-3)))
  \end{lstlisting}
  \caption{\label{fig:pythonpackage}Package definitions of Python and Scikit-learn.}
\end{figure}

Package definitions are declarative and embedded in the Scheme
programming language~\cite{courtes2013:functional}.
\Cref{fig:pythonpackage} shows the definition of the \texttt{python}
package, simplified---we omitted fields that specify dependencies and
build options.  The \texttt{source} field declares an \emph{origin}
indicating that the source of \texttt{python} is the file
\texttt{Python-3.10.7.tar.xz}, a so-called \emph{tarball} to be
downloaded over HTTPS.  Crucially, the \texttt{sha256} field specifies
the cryptographic hash of that file.  When Guix fails to download the
file, or if it obtains a different hash, it immediately aborts---an
obvious prerequisite for reproducible software deployment.

The second package definition in \Cref{fig:pythonpackage} is slightly
different: source code is to be checked out using the Git version
control system (VCS), from the \texttt{1.3.2} tag.  The SHA256 hash, in
this case, is that of the checked out directory once serialized as a
so-called \emph{normalized archive} or \emph{nar}.  The nar format was
initially designed for Nix; unlike the venerable tar format (for ``tape
archive''), it omits Unix metadata unimportant in this context: file
timestamps, access rights, and ownership information.

\begin{figure}
  \begin{tikzpicture}
    \begin{axis}[
        width=1\columnwidth,
        height=0.67\columnwidth,
        table/col sep=comma,
        date coordinates in=x,
        stack plots=y,
        area style,
        cycle list name=cb colors,
        xticklabel style={
          rotate=-67.5,
          anchor=west,
        },
        xtick={
          2019-05-05,
          2019-08-25,
          2019-12-29,
          2020-04-19,
          2020-08-09,
          2020-11-29,
          2021-03-21,
          2021-07-11,
          2021-11-07,
          2022-02-27,
          2022-06-19,
          2022-10-09,
          2023-01-29,
          2023-05-21,
          2023-09-10,
          2024-01-07
        },
        xticklabels={
          {May  5, 2019},
          {Aug. 25, 2019},
          {Dec. 29, 2019},
          {Apr. 19, 2020},
          {Aug.  9, 2020},
          {Nov. 29, 2020},
          {Mar. 21, 2021},
          {Jul. 11, 2021},
          {Nov.  7, 2021},
          {Feb. 27, 2022},
          {Jun. 19, 2022},
          {Oct.  9, 2022},
          {Jan. 29, 2023},
          {May 21, 2023},
          {Sep. 10, 2023},
          {Jan.  7, 2024},
        },
        xticklabel style={font=\footnotesize},
        ymin=0,
        ymax=1,
        xmin=2019-05-05,
        xmax=2024-01-07,
        legend pos=north west,
        legend style={
          cells={anchor=west},
        },
      ]
      \addlegendentry {VCS checkout}
      \addplot table[x=date,y=vcs-rel] {pog-types-high-rel.csv}
               \closedcycle;
      \addlegendentry {File download}
      \addplot table[x=date,y=download-rel] {pog-types-high-rel.csv}
               \closedcycle;
  \end{axis}
  \end{tikzpicture}
  \caption{Relative high-level source types by sampled Guix commit.}
  \label{fig:pog-types-origin}
\end{figure}
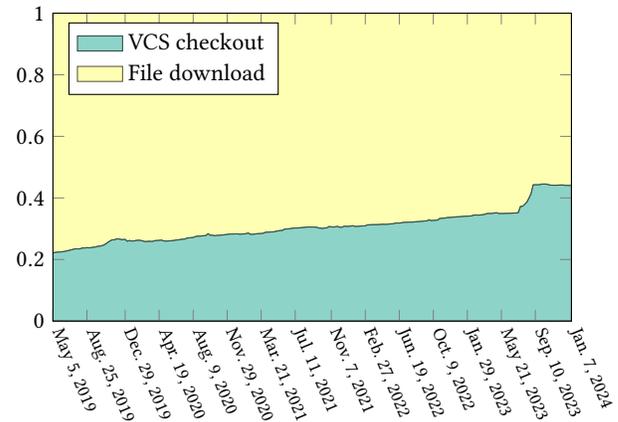

Package source can also be fetched through other methods: by referring
to a Git commit rather than a tag, or by using a different version
control system such as Subversion.  Currently, 44\% of the
packages get their source from a VCS when it was only 22\% back in
2019; \Cref{fig:pog-types-origin} shows how the distribution has changed
over time.
In all these cases, a cryptographic
hash of the content is always specified, as in the examples above.
Source code is thus essentially \emph{content-addressed}, with the URL
and download method serving more as a hint.  An implication is that if
those hints become stale---e.g., the file is no longer available at the
given URL---users can work around the problem: if a copy of the file or
checkout is available elsewhere, one can run the \texttt{guix download}
command with that new URL to feed it to Guix.  Guix then finds the
source with the expected hash and proceeds building it.

Our goal, as designers of a reproducible deployment tool, is to ensure
package source code can always be retrieved automatically and reliably,
even once the original source code hosting site has vanished or has been
compromised.  This is where Software Heritage comes in.

\subsection{Source Code Archiving with Software Heritage}
\label{sec:background:swh}

\emph{Software Heritage}~\cite{dicosmo2017swh} (SWH) is a digital preservation
initiative with the aim of collecting, preserving for the long-term, and
providing access to the entire body of software, in the preferred form for
making modifications to it (referred to as simply ``source code'' in this
article). The \MYURL{https://archive.softwareheritage.org}{SWH
  archive}{2024-01-18} is the largest publicly available archive of software
source code. At the time of writing it contains more than 17 billion unique
source code files and 3.5 billion commits, coming from more than 250 million
projects.

The SWH data model is a deduplicated \emph{Merkle
  graph}~\cite{merkle1987}, with nodes of different types, corresponding to the
software artifacts commonly stored in modern version control systems (VCSs):
individual source code files and directories, commits, releases, etc. SWH hence
actually stores the full development history of software projects, rather than
only the most recent version of archived software products. This enables
restoring a hash-compatible version a Git repository that has disappeared (or
been tampered with) from its usual hosting place---provided it was archived in
time.

Each node in the SWH graph data model can be referenced \textit{via} persistent,
intrinsic identifiers called \emph{SWHIDs}~\cite{dicosmo2018swhid}, which are
computed as Merkle-style SHA1 hashes. For example,\linebreak
\swhid{swh:1:rev:309cf2674ee7a0749978cf8265ab91a60aea0f7d} is a\linebreak SWHID
referencing an archived commit of the Darktable image processing
software, where \texttt{rev}
is the node type: here a ``revision'' (akin to a ``commit'' in Git
parlance).  The current version of
SWHID identifiers (version 1) is compatible with Git SHA1 hash, which
allows revision SWHIDs to be constructed from Git commit identifiers and vice-versa.
(Note that merely constructing a syntactically correct SWHID from Git does not
mean the corresponding Git object has actually been \emph{archived} in SWH.)

The SWH archive is populated primarily by \emph{crawlers} (``pull'' style) that
track major forges (e.g., GitHub, GitLab, \ldots), GNU/Linux distributions
(e.g., Debian), and package manager repositories (e.g., PyPI, NPM,
\ldots). ``Push'' style archiving is also available; particularly relevant for
this work is the \emph{Save Code Now} feature.  It allows users to trigger on-demand
archiving of a specific Git repository that has either never been
archived before or ought to be \emph{re}-archived promptly, before the pull
crawlers have a chance to notice its latest archived copy is out-of-date.

Various mechanisms to access the SWH archive are available, depending on the
use case. Users can browse it \textit{via} a forge-like Web interface at
\url{https://archive.softwareheritage.org}. Developers can integrate with SWH
via various \emph{application programming interfaces} (APIs): Web, gRPC,
file system, and GraphQL-based. We rely on the
\MYURL{https://archive.softwareheritage.org/api/}{Web API}{2024-01-19} for the
Guix/SWH integration described in this paper, using it for verifying that
sources have been archived and triggering push archiving of missing sources.
Researchers can also analyze the SWH archive data in bulk, \textit{via} public
large-scale datasets~\cite{pietri2020swhdataset}.

For the specific need of reconstructing repositories from the archive (e.g., in
case of disappearance from their previous hosting place), a dedicated
asynchronous service called the \emph{Vault} is provided. Its need comes from the
fact that, due to deduplication, an individual repository is stored by SWH as a
(sub-)graph made of many nodes---e.g., tens of millions files, commits, etc.,
for a project like the Linux kernel. A user interested in retrieving a
specific version of the \texttt{linux.git} repository will then file a Vault
``cooking'' request, \textit{via} either the Web user interface or the API, which will then be processed
by a dedicated pool of workers.  When the bundle is ready to be downloaded, the user
is notified, \textit{via} mail or a Webhook trigger.

\section{Implementation}
\label{sec:implementation}

The connection between Guix and Software Heritage goes both ways: first we must
ensure that the SWH archive ingests source code of packages Guix refers to, and
second Guix must fall back to retrieving source code from SWH.  The sections
below look at these two cases.

\begin{figure}
  \includegraphics[width=\linewidth]{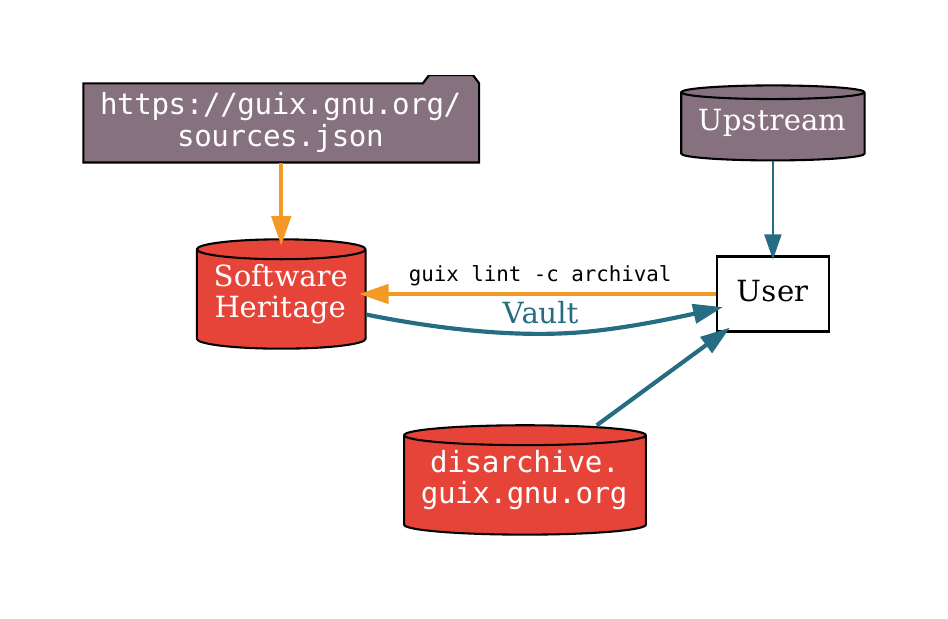}
  \caption{Populating the Software Heritage archive (orange arrows) and
    retrieving source code (blue arrows).}
  \label{fig:flow}
\end{figure}

\subsection{Populating the Archive}
\label{sec:populating}

As we have seen, the SWH archive has already been ingesting source code
from a variety of repositories, covering a large subset of the code
referenced by Guix packages.  Our goal is to achieve \emph{full
coverage}---ensuring that the archive contains all the source code
referenced by Guix packages at any time.  To achieve that, Guix
explicitly triggers source code archiving \textit{via} two mechanisms
represented by the orange arrows in \Cref{fig:flow}.

First, Guix packagers routinely run the \texttt{guix lint} command on
packages they submit for inclusion.  This command checks almost 30
properties on packages, such as making sure that they respect certain
conventions and that the URLs they refer to are reachable.  We extended
\texttt{guix lint} with an \texttt{archival} check that
(1)~checks whether the code is already archived, and (2)~submits a Save
Code Now request (see \Cref{sec:background:swh}) to SWH, \textit{via} the Web
API, if the source is not already archived \emph{and} if it is a VCS
checkout---the Save Code Now interface rejects requests to save
arbitrary tarballs.

The second mechanism complements this: Guix publishes
\url{https://guix.gnu.org/sources.json}, which is a machine-readable
list of all its origins---URLs and hashes, along with commits or
tags when referring to VCS checkouts.  SWH periodically ingests all the
tarballs and VCS repositories referenced by this file using a dedicated
\emph{crawler}, as described in
\Cref{sec:background:swh}.  When it ingests a VCS repository, SWH preserves
all its history; when it ingests a tarball, SWH only preserves the
\emph{contents} of the tarball and not the tarball itself.

\subsection{Retrieving VCS Checkouts}

Ensuring that source code is archived is only half of the job.  How can we
retrieve VCS checkouts from the SWH archive once the original hosting
site has disappeared?

The \texttt{sha256} field of origins in package definitions make them
content-addressed.  However, as we have seen before, Guix and SWH each use a
different method to compute the hash of directories: Guix computes the
hash of a ``normalized archive'' (nar) whereas SWH computes the hash of
a Git tree.  For a long time this mismatch made it impossible to query
content by nar hash---a problem that has now been fixed, as we will see
below.What the SWH archive does permit, though, is to query the checkout
associated with the SHA1 identifier of a Git commit, and to browse the
VCS snapshots of a given URL.  We can distinguish several cases.

The easiest case is that of origins that refer to a Git commit by its
SHA1 identifier.  Using the Web API of the SWH archive, we can
query the \emph{directory} object associated with that identifier---this
is effectively \emph{content-addressed} access, made possible by the
fact that revision identifiers in the SWH data model are equal to Git
commit identifiers.  To obtain the files comprised in that directory,
Guix then uses the SWH Vault; if data has not been ``cooked'' yet, the
download process has to wait until the Vault has made it
available---see the blue arrow on \Cref{fig:flow}.  Our
implementation in Guix is transparent: package definitions do
not need to be changed.  Instead, Guix's download process automatically
falls back to SWH when the URL specified in the origin is unreachable.

What about references to VCS \emph{tags}?  Git tags and commit
identifiers illustrate the trilemma known as ``Zooko's
triangle''~\cite{zooko2021triangle}.  Compared to commit identifiers,
tags have the advantage of being human-readable: the second example of
\Cref{fig:pythonpackage} makes it clear that the intent is to fetch a
checkout for the tag corresponding to version 1.3.2 of Scikit-learn; for
this reason, packagers often use them.  But tags have two major
drawbacks: they are not content-addressed, which make them
context-dependent, and they are mutable---Git allows users to replace a
tag pointing to a given commit with a tag pointing to a different
commit.

In this case, Guix uses the SWH Web API to (1)~look up the repository by
URL, (2)~look up the tag by name to obtain the corresponding commit
identifier, and (3)~download the corresponding directory from the Vault.
We have anecdotal evidence that this process is ``good enough'' in most
cases, but it is inherently brittle and could fail or return the wrong
data: tags might have been modified, URLs might have been reused to host
different code, etc.  The worst that can happen is that Guix is unable
to download the source, although SWH might contain it; hash mismatches are
detected and cause Guix to abort, as we have seen in \Cref{sec:background}.

The fundamental mismatch in how Guix and SWH identify directories was
addressed by a recent SWH feature deployed in January~2024: SWH now
computes and exposes nar hashes for directories.  These hashes are an
extension of the SWH data model called \emph{external identifiers} or
\emph{ExtIDs}; the
\MYURL{https://archive.softwareheritage.org/api/1/extid/doc/}{Web
  API}{2024-01-11} lets us obtain the SWHID corresponding to a
\texttt{nar-sha256} ExtID, which is exactly what is necessary to ensure
content-addressed access in all cases.  Consequently, the fallback code
in Guix was changed to use that method.  Since those ExtIDs have not yet
been computed for previously-ingested origins, Guix still uses the
method described earlier when lookup by \texttt{nar-sha256} fails.

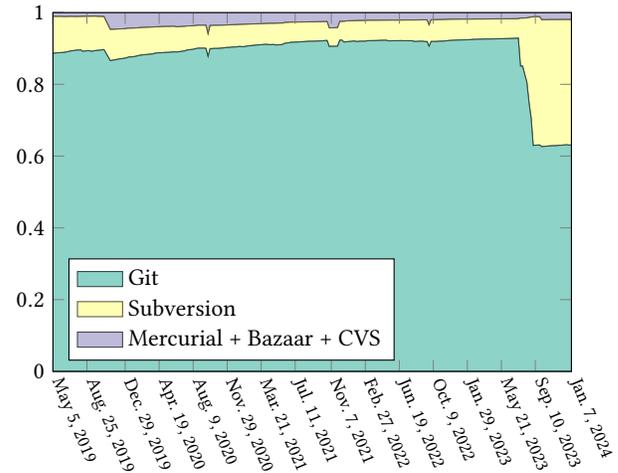
\begin{figure}
  \begin{tikzpicture}
    \begin{axis}[
        width=1\columnwidth,
        height=0.75\columnwidth,
        table/col sep=comma,
        date coordinates in=x,
        stack plots=y,
        area style,
        cycle list name=cb colors,
        xticklabel style={
          rotate=-67.5,
          anchor=west,
        },
        xtick={
          2019-05-05,
          2019-08-25,
          2019-12-29,
          2020-04-19,
          2020-08-09,
          2020-11-29,
          2021-03-21,
          2021-07-11,
          2021-11-07,
          2022-02-27,
          2022-06-19,
          2022-10-09,
          2023-01-29,
          2023-05-21,
          2023-09-10,
          2024-01-07
        },
        xticklabels={
          {May  5, 2019},
          {Aug. 25, 2019},
          {Dec. 29, 2019},
          {Apr. 19, 2020},
          {Aug.  9, 2020},
          {Nov. 29, 2020},
          {Mar. 21, 2021},
          {Jul. 11, 2021},
          {Nov.  7, 2021},
          {Feb. 27, 2022},
          {Jun. 19, 2022},
          {Oct.  9, 2022},
          {Jan. 29, 2023},
          {May 21, 2023},
          {Sep. 10, 2023},
          {Jan.  7, 2024},
        },
        xticklabel style={font=\footnotesize},
        ymin=0,
        ymax=1,
        xmin=2019-05-05,
        xmax=2024-01-07,
        legend pos=south west,
        legend style={
          cells={anchor=west},
        },
      ]
      \addlegendentry {Git}
      \addplot table[x=date,y=git-rel] {pog-types-vcs-rel.csv}
               \closedcycle;
      \addlegendentry {Subversion}
      \addplot table[x=date,y=svn-rel] {pog-types-vcs-rel.csv}
               \closedcycle;
      \addlegendentry {Mercurial + Bazaar + CVS}
      \addplot table[x=date,y=other-rel] {pog-types-vcs-rel.csv}
               \closedcycle;
    \end{axis}
  \end{tikzpicture}
  \caption{Relative VCS source types by sampled Guix commit.}
  \label{fig:pog-types-vcs}
\end{figure}

The beauty of this content-addressed download method is that it works
regardless of the origin type.  In particular, it will still work when
Git repositories start using SHA256 identifiers instead of SHA1---a
feature that is slowly being deployed at this time---and it works for
other version control systems too: Mercurial, Subversion,
and CVS.  Of these, Mercurial and CVS amount for less than 0.2\%
of the sources in the entire package collection, but Subversion amounts
for 15.7\% of all sources as it is used to retrieve the source of the
more than \num{4000} individual \TeX~Live packages---see
\Cref{fig:pog-types-vcs}.

\subsection{Retrieving Source Code Tarballs}
\label{sec:disarchive}

When SWH ingests source code from a tarball (or any other archival file
format), it unpacks the tarball and stores only its contents, rather
than keeping the entire file.  This is the natural approach given its
graph-oriented representation of software projects.  On the other hand,
Guix considers the tarball an atomic input and expects to be able to
retrieve it intact and verify that it is unmodified.  The process of
creating a tarball is, in general, not reproducible: even when using the
same inputs and tools, timestamps and nondeterminism can result in small
differences in the resulting files~\cite{lamb2021:reproducible}.
Therefore, Guix cannot directly retrieve a tarball from SWH, nor can it
retrieve the same contents and synthesize one that would pass its own
verification.

To solve this mismatch, we have developed a way to decompose tarballs
into two parts: the file system content and a description of the process
that transformed it into the tarball.  If the description is
sufficiently detailed, we can run this process in reverse and arrive
again at the original tarball given these two parts.  Since SWH already
stores the contents, if we could provide the corresponding description,
Guix could revive the original tarball.

\begin{figure}
  \begin{tikzpicture}[
        crucial/.style = {
          text width=17mm, minimum height=15mm,
          text centered, rounded corners,
          fill=white, text=black,
          draw=guixorange1, line width=1mm
        },
        input/.style = {
          shape=circle, text centered, text width=12mm,
          fill=white, text=black,
          draw=black, very thick
        },
        important/.style = {
          minimum height=12mm, text width=35mm,
          text centered, rounded corners,
          fill=white, text=black
        }]
    \matrix[row sep=3mm, column sep=3mm] {
      & \node(swh) [important] {file content at SWH \texttt{swh:1:dir:cabba93}\ldots}; &
      \\
      \node(disarchiveout) [crucial, draw=guixblue1] {Disarchive \\[1mm] assemble};
      & \node(tarball) [input] {tarball \texttt{tar.gz}};
      & \node(disarchivein) [crucial] {Disarchive \\[1mm] disassemble};
      \\
      & \node(database) [important] {tarball metadata \texttt{disarchive.guix.gnu.org}};
      &
      \\
    };

    \path[very thick, draw=guixorange1] (tarball) edge [->] (disarchivein);
    \path[very thick, draw=guixorange1] (disarchivein) edge [->, out=90, in=0] (swh);
    \path[very thick, draw=guixorange1] (disarchivein) edge [->, out=-90, in=0] (database);

    \path[very thick, draw=guixblue1] (swh) edge [->, in=90, out=180] (disarchiveout);
    \path[very thick, draw=guixblue1] (database) edge [->, out=180, in=-90] (disarchiveout);
    \path[very thick, draw=guixblue1] (disarchiveout) edge [->] (tarball);
  \end{tikzpicture}

  \caption{Disarchive tarball disassembly (orange arrows) takes a ``tarball'' as input
    and produces metadata along with a SWHID pointing to the tarball
    contents.  Assembly (blue arrows) reconstructs the tarball by combining its metadata
    and its contents.}
  \label{fig:disarchive}
\end{figure}
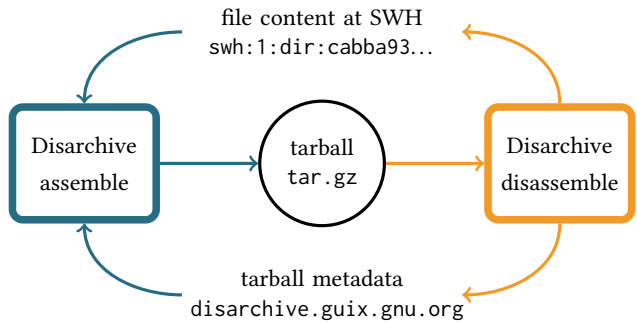

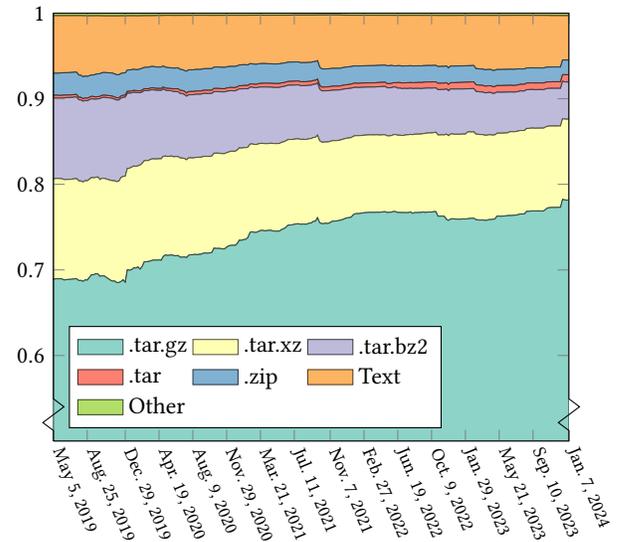
\begin{figure}
  \begin{tikzpicture}
    \begin{axis}[
        table/col sep=comma,
        date coordinates in=x,
        stack plots=y,
        area style,
        cycle list name=cb colors,
        xticklabel style={
          rotate=-67.5,
          anchor=west,
        },
        xtick={
          2019-05-05,
          2019-08-25,
          2019-12-29,
          2020-04-19,
          2020-08-09,
          2020-11-29,
          2021-03-21,
          2021-07-11,
          2021-11-07,
          2022-02-27,
          2022-06-19,
          2022-10-09,
          2023-01-29,
          2023-05-21,
          2023-09-10,
          2024-01-07
        },
        xticklabels={
          {May  5, 2019},
          {Aug. 25, 2019},
          {Dec. 29, 2019},
          {Apr. 19, 2020},
          {Aug.  9, 2020},
          {Nov. 29, 2020},
          {Mar. 21, 2021},
          {Jul. 11, 2021},
          {Nov.  7, 2021},
          {Feb. 27, 2022},
          {Jun. 19, 2022},
          {Oct.  9, 2022},
          {Jan. 29, 2023},
          {May 21, 2023},
          {Sep. 10, 2023},
          {Jan.  7, 2024},
        },
        xticklabel style={font=\footnotesize},
        ymin=0.5,
        ymax=1,
        axis y discontinuity=crunch,
        ytick={0.6,0.7,0.8,0.9,1},
        xmin=2019-05-05,
        xmax=2024-01-07,
        legend pos=south west,
        legend columns=3,
        legend style={
          cells={anchor=west},
        },
      ]
      \addlegendentry {.tar.gz}
      \addplot table[x=date,y=tar-gz-rel] {pog-types-dl-rel.csv}
               \closedcycle;
      \addlegendentry {.tar.xz}
      \addplot table[x=date,y=tar-xz-rel] {pog-types-dl-rel.csv}
               \closedcycle;
      \addlegendentry {.tar.bz2}
      \addplot table[x=date,y=tar-bz2-rel] {pog-types-dl-rel.csv}
               \closedcycle;
      \addlegendentry {.tar}
      \addplot table[x=date,y=tar-rel] {pog-types-dl-rel.csv}
               \closedcycle;
      \addlegendentry {.zip}
      \addplot table[x=date,y=zip-rel] {pog-types-dl-rel.csv}
               \closedcycle;
      \addlegendentry {Text}
      \addplot table[x=date,y=text-rel] {pog-types-dl-rel.csv}
               \closedcycle;
      \addlegendentry {Other}
      \addplot table[x=date,y=other-rel] {pog-types-dl-rel.csv}
               \closedcycle;
    \end{axis}
  \end{tikzpicture}
  \caption{Relative download source types by sampled Guix commit
    (truncated at 50\%).}
  \label{fig:pog-types-download}
\end{figure}

We created a tool called
\MYURL{https://ngyro.com/software/disarchive.html}{Disarchive}{2024-01-19}
that implements this strategy: it can \emph{disassemble} a tarball to
obtain a description and a link to its contents in SWH,
and can \emph{assemble} that tarball
given its contents and the description, as shown in
\Cref{fig:disarchive}.  Currently, Disarchive supports
decomposing plain tar files as well as compressed files in gzip, bzip2,
and XZ formats.  These formats represent the majority of non-VCS Guix
sources (see \Cref{fig:pog-types-download}).  For tar files, it stores
the metadata for each file in
the order the files appeared in the original tarball: this includes file
metadata (modification time, Unix owner and group, permissions) but also
low-level details about the tar headers (details in the representation
of tar members that vary between tar implementations and variants).

To describe a compressed file, Disarchive guesses the compression system
used and then verifies its guess---e.g., it tries both GNU gzip and
zlib, with various flags, until it matches the given gzip file.  This
brute-force approach is admittedly not very elegant, but it works in
practice.  Testing shows that 97.3\% of the 41,521 compressed tarballs
referred to by Guix (see \Cref{sec:pog}) can be disassembled using this
approach.

\begin{figure}
  \begin{small}
\begin{lstlisting}[language=Scheme, escapeinside={||}]
(disarchive
  (version 0)
  (gzip-member
    (name "sed-4.8.tar.gz")
    (digest (sha256 "53cf3e1|\ldots{}|"))
    (header (mtime 0) (extra-flags 2) (os 3))
    (footer (crc 1582442600) (isize 10516480))
    (compressor gnu-best-rsync)
    (input
      (tarball
        (name "sed-4.8.tar")
        (digest (sha256 "626c2e3|\ldots{}|"))
        (default-header
          (chksum (trailer " "))
          (typeflag 0)
          (magic "\x00\x00\x00\x00\x00\x00")
          (version "\x00\x00")
          (data-padding ""))
        (headers
          ("sed-4.8/"
           (mode 493)
           (mtime 1579061438)
           (chksum 3662)
           (typeflag 53))
          ;; many headers omitted
          ("sed-4.8/bootstrap.conf"
           (size 3129)
           (mtime 1578639009)
           (chksum 5071)))
        (padding 1024)
        (input
          (directory-ref
            (version 0)
            (name "sed-4.8")
            (addresses (swhid "swh:1:dir:f36d96f|\ldots{}|"))
            (digest (sha256 "994ba02|\ldots{}|"))))))))
\end{lstlisting}
  \end{small}
  \caption{Disarchive disassemble output for \texttt{sed-4.8.tar.gz}.\label{fig:disarchivedescription}}
\end{figure}

The tarball description resulting from Disarchive's disassemble step
looks like that shown in \Cref{fig:disarchivedescription}.  It is a tree
(a Scheme s-expression) showing, in this case, the components of the
gzipped tar file: there is first a \emph{gzip member} with a header
and footer and whose body was compressed with the
\texttt{gnu-best-rsync} method; the gzip member has a \emph{tarball} as
its input, which, in turn, has a \emph{directory reference} as its
input.  That last element points to the actual contents of the tarball,
referred to by a SWHID.

As can be guessed from this example, the difficult part in designing
Disarchive was to find out which details about tar and compression
formats needed to be kept so that Disarchive could faithfully represent
the variety of tarballs actually used without loss of information while
keeping the output reasonably compact.  The format shown here has proved
to meet these requirements for many thousands of tarballs, as we will
see in \Cref{sec:evaluation}.  The storage costs for those descriptions
are modest.  As an example, for \texttt{sed-4.8.tar.gz}, which contains
987 files, the description takes 140~KiB uncompressed and only 14~KiB
when compressed with gzip at its maximum level.

Using Disarchive, we have built and are continuously updating a
\MYURL{https://disarchive.guix.gnu.org}{database}{2024-01-19} of these
tarball descriptions.  If a source tarball is no longer available from
its original location, Guix can automatically recreate it with
Disarchive by downloading its contents from SWH and its description
from the database---see the second blue arrow on \Cref{fig:flow}.

Maintaining a separate database like this is at odds with our stated
goal of full coverage by the SWH archive.  As such, work is ongoing to
integrate Disarchive and its database into SWH properly, so that these
descriptions can be archived there for the long term.  Doing so will not
be particularly costly in terms of archival storage: Disarchive
descriptions are really small in comparison to other source code
artifacts already archived by SWH.  In conjunction with ExtIDs, this
will enable the SWH archive to produce tarballs that are
byte-identical to tarballs observed in the wild---e.g., as a new Vault
download format.

\subsection{Limitations and Mitigation}
\label{sec:limitations}

Our implementation suffers from several limitations, as we have seen
before.  Source code tarballs are the most challenging part.  As of this
writing, some archive formats are not supported by Disarchive, including
lzip compression, Zip files, and some unusual forms of gzip compression.
As we will see in \Cref{sec:evaluation}, these amount to less than
2.6\% of the package source code though.

The Subversion version control system poses unique challenges: unlike
other VCSes, it allows users to retrieve individual sub-directories
within a repository.  Luckily, SWH now computes \texttt{nar-sha256}
ExtIDs for them, allowing Guix to recover them.  However, \TeX~Live
packages in Guix work by \emph{combining} several directories checked
out from Subversion---typically a source and a documentation
directory---and the nar hash in package definitions is computed over
that combination.  This defeats content-addressed lookups because those
combinations do not exist as such in the SWH archive.  As of this
writing, the Guix and SWH team are discussing a solution whereby SWH
would also store the nar hash for these combinations.

Another limitation has to do with the way Guix uses the SWH Vault: the
Vault needs to ``cook'' source code archives before Guix can download
them, and that process can take from minutes to days depending on the
size of the artifact to be built and the load of the service.  That we
cannot guarantee timely downloads is a significant usability problem.
One solution we are considering is to have SWH ``pre-cook'' any source
referenced by Guix packages.  The storage cost of such a policy change
is currently being assessed before implementation and deployment can
proceed.

\section{Evaluation}
\label{sec:evaluation}

In this section we evaluate ``link rot''---source code disappearance
over time---and whether the upstream source code of Guix packages
exists in the SWH archive.

\subsection{Source Code ``Link Rot''}

The first question we asked ourselves is: what is the extent of the
problem?  How much of the source code referenced by Guix packages has
disappeared or has been tampered with?  To answer this question, we
attempted to download the source code of all the packages found in Guix
at the time of past releases, covering 5 years of history from version
1.0.0 (released in May 2019) to today (January 2024).  Even though Guix
was already 6 years old when 1.0.0 was released, we choose this as the
starting point because it represents the oldest point in history
supported as a target for \texttt{guix time-machine}.

\begin{table}
  \footnotesize
  \caption{``Link rot'' empirical evaluation of all package sources over
    five years.}
  \label{tab:link-rot}
  \begin{tabular}{@{}lrrrrr@{}}
    \toprule
    &\scriptsize{May 2019}&\scriptsize{Apr. 2020}&\scriptsize{Nov. 2020}&\scriptsize{May 2021}&\scriptsize{Dec. 2022}
    \\
    &v1.0.0&v1.1.0&v1.2.0&v1.3.0&v1.4.0
    \\ \midrule
    \#sources&8794&11659&13609&15520&20184
    \\ \midrule
    avail.&91.5\%&92.4\%&95.0\%&95.7\%&96.4\%
    \\
    missing&8.5\%&7.6\%&5.0\%&4.3\%&3.6\%
    \\
    hash mis.&87&63&69&66&52 \\
    \bottomrule
  \end{tabular}
\end{table}

To estimate link rot, we re-downloaded all the source code of all the
packages defined in Guix since version 1.0.0 from their upstream
location, turning off the fallback mechanisms described in
\Cref{sec:implementation}.  \Cref{tab:link-rot} shows the fraction of
package source code that could still be downloaded from its initial
location.  After one year, 96.4\% of the package source code is still
available unaltered from its upstream location; it decreases to 91.5\%
after five years.  Among the sources reported as \textit{missing}, a
fraction was actually still available for download but had been tampered
with (\textit{hash mismatch}): that represents 1\% of the sources after
five years, and 0.3\% after one year.  These findings are consistent
with those reported in an earlier
study~\cite{escamilla2023:swh-coverage}.

To put it in perspective, 3.6\% corresponds to 726 packages of version
1.4.0 that are already ``lost'' a year later.  How serious is this?
Obviously, this largely depends on what packages one is trying to
deploy, directly but also \emph{indirectly}.  Looking at the number of
dependents of the packages whose source is missing (their \textit{rank}
in the graph) gives a more accurate picture.  As an example,
\texttt{openjdk-9.181.tar.bz2} is unavailable from its original upstream
URL as it appears in Guix~1.4.0; the \texttt{openjdk 9.181} package had
184 dependents, so we would effectively be losing \emph{185 packages},
not just one, if this tarball were unrecoverable.  Merely looking at the
fraction of missing package sources underestimates the real impact.

\subsection{Preservation of Guix}
\label{sec:pog}

The second question we wanted to answer is this: how much of the source
code of Guix packages is actually archived in SWH?  Note that we are
concerned here with \emph{preservation}.  Code available in SWH is not
necessarily \emph{recoverable} today by Guix due to the limitations
outlined in \Cref{sec:limitations}.  Availability is a necessary
condition for code to be recoverable, though.

To answer this question, we implemented a set of tools on top of Guix to
conduct the analyses described below~\cite{preservation-of-guix}.
We have extracted a list of nearly all of the sources
referenced by Guix since version 1.0.0.  We sampled the
history of the Guix repository, analyzing one commit per week from May
2019~\cite{guix2019} until January 2024~\cite{guix2024}.\footnote{Sampling
reduces the number of commits
to analyze by tens of thousands, easing the computational burden of
analysis.  However, any external references that appeared and
disappeared between samples are missed.  The number of missed sources
can be estimated by searching the text of the Guix repository (over the
given period) for cryptographic hashes.  Such a search suggests that
only about \num{1500} (2.1\%) sources were missed due to sampling.}  This
gives us 243 snapshots of the history of the package collection.
At each snapshot the package graph was built and crawled, and each
external reference was collected as both a cryptographic hash and a
machine-readable description of how to obtain it.  There are \num{68473}
sources in total.

As described in \Cref{sec:implementation}, the cryptographic hashes from
Guix cannot generally be used to locate content in the SWH archive.
Therefore, we attempted to download and verify these sources in order to
compute SWHIDs for each of them.
Tarball sources are processed with Disarchive,
and the underlying directory SWHID is extracted from the Disarchive
specification.  For bare files such as patches, we compute a content
SWHID.  Git sources are cloned and the directory SWHID of the checkout
is taken from Git itself.  Subversion sources in Guix can be composed of
selected sub-directories of a checkout.  To account for this, we
take the SWHID of each subdirectory.  Other, smaller categories of
sources, like Mercurial and CVS, are ignored.

This approach can fail in a number of ways: the sources may be already
unavailable; they may be available but fail to verify because they have
changed since they were first referenced by Guix; or Disarchive may be
unable to process a tarball.  Nevertheless, we found SWHIDs for \num{64480}
(94.2\%) of the sources.

\begin{table}
  \footnotesize
  \caption{SWH archive coverage of collected sources, including coverage of select commits.}
  \label{tab:pog}
  \clearpage{}\begin{tabular}{@{}llrrrrr@{}}
  \toprule
  \textbf{Commit}
    & \textbf{Date}
    & \multicolumn{2}{c}{\textbf{Stored}}
    & \multicolumn{2}{c}{\textbf{Missing}}
    & \textbf{Total} \\
  \midrule
  \texttt{ee3ce0d} & mag   5, 2019 & \num{6313} & 71.7\% & \num{1879} & 21.3\% & \num{8810} \\
  \texttt{3d76112} & ago. 25, 2019 & \num{6870} & 73.4\% & \num{1812} & 19.4\% & \num{9362} \\
  \texttt{34085ea} & dic. 29, 2019 & \num{8146} & 79.1\% & \num{1524} & 14.8\% & \num{10300} \\
  \texttt{fafe234} & apr. 19, 2020 & \num{9470} & 79.8\% & \num{1453} & 12.2\% & \num{11863} \\
  \texttt{cb97d07} & ago.  9, 2020 & \num{11430} & 87.5\% & \num{613} & 04.7\% & \num{13070} \\
  \texttt{60a10a1} & nov. 29, 2020 & \num{12388} & 88.7\% & \num{507} & 03.6\% & \num{13963} \\
  \texttt{ba0dc1d} & mar. 21, 2021 & \num{13894} & 89.4\% & \num{443} & 02.9\% & \num{15541} \\
  \texttt{5c3489a} & lug. 11, 2021 & \num{15100} & 90.6\% & \num{353} & 02.1\% & \num{16666} \\
  \texttt{b11badf} & nov.  7, 2021 & \num{16247} & 90.9\% & \num{306} & 01.7\% & \num{17879} \\
  \texttt{31ecd80} & feb. 27, 2022 & \num{17470} & 91.9\% & \num{274} & 01.4\% & \num{19016} \\
  \texttt{77db24f} & giu. 19, 2022 & \num{17985} & 92.1\% & \num{275} & 01.4\% & \num{19536} \\
  \texttt{79358a9} & ott.  9, 2022 & \num{18805} & 92.3\% & \num{271} & 01.3\% & \num{20373} \\
  \texttt{bea2240} & gen. 29, 2023 & \num{19367} & 92.9\% & \num{278} & 01.3\% & \num{20852} \\
  \texttt{7b3f571} & mag  21, 2023 & \num{20323} & 92.8\% & \num{326} & 01.5\% & \num{21903} \\
  \texttt{2eb6df5} & set. 10, 2023 & \num{24775} & 94.0\% & \num{378} & 01.4\% & \num{26370} \\
  \texttt{25bcf4e} & gen.  7, 2024 & \num{25473} & 93.8\% & \num{349} & 01.3\% & \num{27157} \\
  \cmidrule(r){1-2}
  \multicolumn{2}{c}{\emph{All commits}}
    & \num{58530} & 85.5\%
    & \num{5950} & 08.7\%
    & \num{68473} \\
  \bottomrule
\end{tabular}
\clearpage{}
\end{table}

The SWH Web API provides the
\MYURL{https://archive.softwareheritage.org/api/1/known/doc/}{\texttt{known}}{2024-02-08}
endpoint to query the existence of SWHIDs in bulk.  Using this, we
found that the SWH archive covered \num{58530} (85.5\%) of the total
sources, or 90.8\% of the sources for which we found SWHIDs---see
\Cref{tab:pog}.

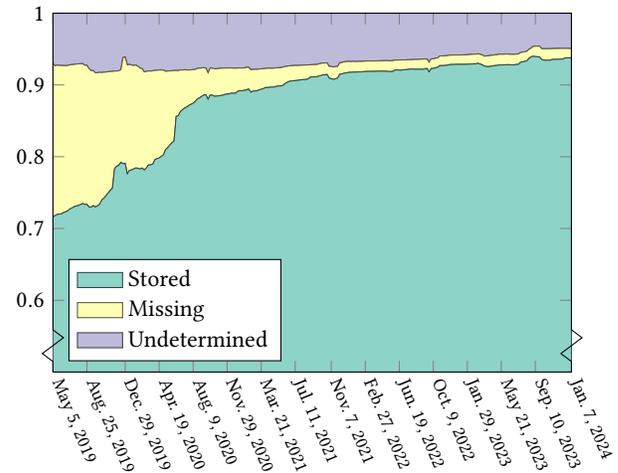
\begin{figure}
  \begin{tikzpicture}
    \begin{axis}[
        width=1\columnwidth,
        height=0.75\columnwidth,
        table/col sep=comma,
        date coordinates in=x,
        stack plots=y,
        area style,
        cycle list name=cb colors,
        xticklabel style={
          rotate=-67.5,
          anchor=west,
        },
        xtick={
          2019-05-05,
          2019-08-25,
          2019-12-29,
          2020-04-19,
          2020-08-09,
          2020-11-29,
          2021-03-21,
          2021-07-11,
          2021-11-07,
          2022-02-27,
          2022-06-19,
          2022-10-09,
          2023-01-29,
          2023-05-21,
          2023-09-10,
          2024-01-07
        },
        xticklabels={
          {May  5, 2019},
          {Aug. 25, 2019},
          {Dec. 29, 2019},
          {Apr. 19, 2020},
          {Aug.  9, 2020},
          {Nov. 29, 2020},
          {Mar. 21, 2021},
          {Jul. 11, 2021},
          {Nov.  7, 2021},
          {Feb. 27, 2022},
          {Jun. 19, 2022},
          {Oct.  9, 2022},
          {Jan. 29, 2023},
          {May 21, 2023},
          {Sep. 10, 2023},
          {Jan.  7, 2024},
        },
        xticklabel style={font=\footnotesize},
        ymin=0.5,
        ymax=1,
        axis y discontinuity=crunch,
        ytick={0.6,0.7,0.8,0.9,1},
        xmin=2019-05-05,
        xmax=2024-01-07,
        legend pos=south west,
        legend style={
          cells={anchor=west},
        },
      ]
      \addlegendentry {Stored}
      \addplot table[x=date,y=stored-rel] {pog-status-rel.csv}
               \closedcycle;
      \addlegendentry {Missing}
      \addplot table[x=date,y=missing-rel] {pog-status-rel.csv}
               \closedcycle;
      \addlegendentry {Undetermined}
      \addplot table[x=date,y=unknown-rel] {pog-status-rel.csv}
               \closedcycle;
    \end{axis}
  \end{tikzpicture}
  \caption{Relative SWH archive coverage by sampled Guix commit (truncated at 50\%).}
  \label{fig:pog}
\end{figure}

Breaking down the results by commit shows that recent commits have far
better coverage than older commits.  The earliest commit is missing
21.3\% of total sources while the latest commit is missing only
1.3\%.  This difference is clear in \Cref{fig:pog}, which shows relative
SWH coverage by the publish date of individual commits.  The main reason
for this improvement is that SWH started loading sources as listed by
Guix in September 2020 (see \Cref{sec:populating}).  The sources
collected for this report will be loaded by SWH using the same system,
which will improve the coverage of earlier commits.

The true coverage is likely even better than these numbers show, as many
of the sources we were unable to process may be in the SWH archive.
This suggests that \emph{complete preservation} is an achievable goal.

\subsection{Automatic Source Code Recovery}

It is one thing to know that source code is \emph{preserved} by SWH; it
is a different thing to be able to automatically \emph{recover} it.  This
is in part due to the limitations discussed in \Cref{sec:limitations}
but also, more importantly, due to the fact that the recovery mechanism
\emph{is itself} improving over time.  Consider this command:

\begin{verbatim}
guix time-machine -q --commit=v1.0.0 -- install emacs
\end{verbatim}

It installs (and potentially rebuilds) Emacs and all its dependencies as
they were defined in Guix 1.0.0 from 2019.  SWH support back then was in
its infancy: it was able to retrieve Git checkouts and little more.
Thus, the command above ends up using this less-capable code, or even
buggy code, which may fail to recover source code, even though today's
SWH support in Guix can do so.  Two bugs illustrate the problem: the
fallback mechanism currently
\MYURL{https://issues.guix.gnu.org/28659}{does not fire upon hash
  mismatches}{2024-02-11}, meaning that source that has been tampered
with upstream is \emph{not} automatically recovered; the Vault Web API
recently started responding with HTTP redirects that
\MYURL{https://issues.guix.gnu.org/69058}{Guix code did not
  follow}{2024-02-12}, preventing automatic recovery altogether.  These
are ``normal'' bugs that happen in any software development effort and
eventually get fixed, but because Guix lets users travel back in its
history, those bugs make \emph{automatic} recovery a real challenge.

One measure we took recently to mitigate that is to further
\emph{decouple} the download mechanism from the rest of the packaging
machinery.  For example, downloads of files as well as Git checkouts can
now be delegated to the Guix build daemon (\emph{via} the special
\texttt{builtin:download} and \texttt{builtin:git-download}
``builders'').  The build daemon evolves and incorporates improvements
in its fallback code; those improvements will be beneficial even when
downloading today's source code 5 years from now.  A second mitigation
on our road map is a new \texttt{guix} command to recover source code
referenced by past revisions using present-day techniques.

\section{Related work}
\label{sec:related}

More and more frequently, scientists willing to share their
computational workflows and to make them reproducible resort to
\emph{workflow systems} such as Snakemake and Nextflow.  These tools, in
turn, often delegate deployment of the computational environment to
\emph{container engines} such as Docker, podman, or
Singularity/Apptainer~\cite{grayson2023:workflows}.

Container engines have been advocated as a tool for reproducible
research workflows for almost a decade now~\cite{boettiger2015:docker}.
Container engines run images typically built from a large base
image on top of which additional software is installed by various
means---using the distribution's package manager or additional tools
such as Conda and pip.  By shipping the container image, one enables
others to re-run the exact same code.  Unfortunately, the build process
of those images is rarely reproducible, and provenance tracking and the
source/binary correspondence are lost~\cite{vallet2022:toward}.  As a
result, recipients cannot really tell what software they are running,
nor experiment with variants of that software.

Maneage~\cite{akhlaghi2021:maneage} is a framework that aims to support
reproducible computational workflows and reproducible paper authoring.
To do so, it provides a set of Makefiles to download and build software
from source.  However, its download method has no fallback: one can no
longer deploy the workflow, potentially irreversibly, once one of the
source code tarballs has become unavailable.

Many free operating system distributions and deployment tools predate
Guix of course, and they all have had to ensure to some level that
source code of their packages is available.  However, few have a stated
goal of allowing ``time travel''---being able to redeploy and possibly
rebuild software at a later point in time~\cite{legrand2023:rescience}.  Most major distributions
such as Gentoo, Debian, and FreeBSD Ports maintain copies of the source
code of their packages, at least for a certain amount of time. Most
notably Debian operates the \url{https://snapshot.debian.org} service,
which contains all version of all Debian packages in both source and
binary form (including development versions never shipped in a stable
release), but ``only'' going as far back as 2005 (Debian was created in
1993).  Older releases of Debian are available \textit{via} the regular
distribution network, but at the coarser granularity of stable
releases.

Conversely, many of the package managers that fill a particular niche,
such as Brew, Conda, pip, or Spack, do not have source code mirrors in
place.  Package definitions in Brew and Spack, for example, refer
directly to the upstream source code location, with limited mitigation
against disappearance or tampering of source code.  The pip package
manager fetches packages from the Python Package Index (PyPI), where
some packages are available as binaries only, in the ``wheels'' format;
remaining packages are available as source and retained for an
indefinite amount of time though the project does not commit to any
retention policy.  Spack implements a
\MYURL{https://spack.readthedocs.io/en/latest/mirrors.html}{source code
  mirror mechanism}{2024-02-06} but the project does not appear to
maintain such a cache.  Brew and Conda do not have any source code
mirroring or download fallback in place, to our knowledge.

The deployment model pioneered by Nix and that Guix builds upon gives a
natural solution to source code caching~\cite{dolstra2004nix,
  courtes2022:secure}.  The mechanism that allows users to download
pre-built binaries as a \emph{substitute} for a local build also
applies to source code.  Consequently, substitute servers automatically
act as a cache for source code too.  This solution, however, typically
does not offer durability guarantees: those cached sources may end up
being deleted from substitute servers after an unspecified amount of
time.

The mechanism described in \Cref{sec:populating} that allows SWH to
pull a list of source code URLs from a JSON file published by the Guix
project was initially implemented by a Nix developer.  The Nix project
publishes a JSON file similar to that of Guix, which thus allows Nix to
ensure that its own package source code is also being archived.
However, to date, Nix and its package collection do not implement a
SWH-based download fallback as described in \Cref{sec:implementation}.

SwhFS~\cite{allancon2021swhfs} is a file system for Linux
that allows users to ``mount'' subsets of the Software Heritage archive and
navigate them as it they were part of the local file system. It is meant
to bridge the gap between common software development activities such
as searching through development history and source code archiving. As
such, it is possible to use SwhFS to retrieve code disappeared
from its previously known hosting place, \textit{via} the relevant SWHID
identifiers. It is not possible to do so \textit{via} tarball hash though,
as the SwhFS interface is based on SWHIDs. The persistence
characteristics of SwhFS are inherited from Software Heritage, and
hence analogous to those of the approach proposed in this paper.

Disarchive, described in \Cref{sec:disarchive}, is, to our knowledge, a
new approach to the tarball archiving issue.  In 2007, the Debian Project
developed the
\MYURL{https://manpages.debian.org/unstable/pristine-tar/pristine-tar.1.en.html}{pristine-tar
  tool}{2024-02-06} to address this problem.  To preserve a compressed
file, pristine-tar uses a similar guessing approach to Disarchive, and
was its inspiration.  To preserve a tarball, it generates a fresh
tarball in a controlled way using the system implementation of
\texttt{tar}, and then stores a \emph{binary delta} between the fresh
tarball and the original~\cite{debian2023:pristine-tar}.  Later it can
generate another fresh tarball and apply that delta to recreate an exact
copy of the original upstream tarball.  Disarchive instead opts for a
more transparent approach with an explicit representation of tarball
metadata.  This avoids potential stability issues with the output of the
\texttt{tar} program, and on tests of a few thousand tarballs, results
in smaller descriptions on average.

\section{Conclusion}
\label{sec:conclusion}

The ability to \emph{preserve} source code in a long-term archive and to
automatically \emph{recover it} from the archive when deploying software
might sometimes be dismissed as a technicality.  Our view is that addressing this
technical issue is in fact the very first step required to achieve
reproducible software deployment, itself a prerequisite of reproducible
research workflows.

By connecting Guix to the Software Heritage archive and by designing
Disarchive to bridge the gap between them, we have been able to ensure
that 94\% of the packages provided by Guix today have their source code
archived---85\% if we look at all the packages that have been provided by Guix
over the past 5 years.  Our experience is that automatic source code
recovery is even more challenging than preservation; future work
includes allowing users to recover old package source code using
present-day techniques.
To our knowledge, this is the first time a
software deployment tool is backed by a stable archive.  The relevance
of this work goes beyond Guix itself and extends to the wealth of
deployment tools and scientific workflow managers that similarly need to
be able to reliably download package source code.

Our ambition with Guix is to provide a tool that can redeploy the same
software environment years later, whether or not pre-built binaries are
available---a tool scientists can rely on to share their work among
peers and to inspect and modify it.  The work presented here naturally
stems from this goal---and so do reproducible builds and the functional
deployment model embraced by Guix.  It is at odds with the more
widespread approach that consists in storing pre-built software images
with little or no provenance tracking and little or no support for
inspection and experimentation.

Ten years of experience with Guix have allowed us to identify further
challenges towards that goal---addressing problems that do not appear
when ``time-traveling'' at the scale of one year, such as the influence
of system time and hardware details on software build processes.  Our
work, going forward, is to tackle those remaining stumbling blocks to
achieve reproducible software deployment viable over the course of many
years.

\section{Data Availability}
\label{sec:data-availability}

A full reproducibility package for this work is available for download from Zenodo at \url{https://doi.org/10.5281/zenodo.11256698}.

\clearpage
\balance


\end{document}